\documentclass[11pt, aps, prd, tightenlines,nofootinbib, superscriptaddress,notitlepage]{revtex4-1}
\usepackage{xspace}
\usepackage[english]{babel}
\usepackage{amsfonts}
\usepackage{amsmath, amsthm, amssymb, mathrsfs}
\usepackage{graphicx}
\usepackage[left=2cm, right=2cm, top=2.5cm, bottom=2cm]{geometry}
\usepackage{bbm}
\usepackage[colorlinks=true, citecolor=cyan, linkcolor=., urlcolor=.]{hyperref}
\usepackage{enumerate}
\usepackage{color}
\usepackage{soul}
\usepackage{epsfig}
\usepackage{pst-grad}
\usepackage{pst-plot}

\newcommand{\abs}[1]{\left|#1\right|}

\newcommand{\ket}[1]{|#1\rangle}
\newcommand{\norm}[1]{\left|\left|#1\right|\right|}

\def\eqref#1{(\textcolor{blue}{\ref{#1}})}

\begin{document}

\title{Graph coherent states for loop quantum gravity}

\author{Mehdi Assanioussi}
\email[]{mehdi.assanioussi@fuw.edu.pl}
\email[]{mehdi.assanioussi@desy.de}
\affiliation{Faculty of Physics, University of Warsaw, Pasteura 5, 02-093 Warsaw, Poland.}
\affiliation{II. Institute for Theoretical Physics, University of Hamburg,\\ Luruper Chaussee 149, 22761 Hamburg, Germany.}

\begin{abstract}
In this article we further investigate the construction of graph coherent states, first introduced in \cite{GCS1}, in the context of loop quantum gravity. We specifically investigate the possibility of defining a family of graph coherent states adapted to the canonical loop quantum gravity Hamiltonian. After discussing various aspects of the general framework and the choice of operators, we use the Euclidean part of the Hamiltonian operator to propose a generator of the generalized canonical structure, necessary to define the coherent states. We then apply the construction procedure, leading to a new family of graph coherent states partially adapted to the gravity Hamiltonian in loop quantum gravity.
\end{abstract}

\maketitle

\section{Introduction}\label{sec1}

In a previous article \cite{GCS1}, a new class of coherent states has been introduced in the context of loop quantum gravity (LQG) \cite{LQG0,LQG1,LQG2,LQG3,LQG4,LQG5}: the graph coherent states. These are canonical coherent states with respect to a generalized canonical structure which generates a graph change on the spin networks graphs. The graph change is seen as an excitation which distinguishes the spin networks graphs, and induces a decomposition of the loop Hilbert space into separable subspaces, each characterized by an ordered family of graphs. Naturally, the graph coherent states take the form of a normalized infinite superposition of spin network states, each being labeled by the number of excitations they carry and a canonical vacuum state. The goal of introducing such states is to improve the analysis and understanding of graph changing operators in loop quantum theories, such as the Hamiltonian operators.

The construction we introduced in \cite{GCS1} was general in the sense that it may be applied to various graph changes and in the context of a loop quantum theory with arbitrary compact gauge group. In particular, there is a large freedom in choosing the mapping between the intertwiners in the case of a non Abelian gauge theory. This freedom is crucial as it allows to adapt the choice of graph coherent states to a particular family of operators of interest. We furthermore constructed a concrete example, using the ``special loop'' graph change, which is compatible with Yang-Mills Hamiltonian operator, and we established some coherence properties of the induced graph coherent states in that case. In the current article, we focus on the gravity Hamiltonian which has a more complex structure than the one of Yang-Mills Hamiltonian. Namely, while considering roughly the same graph change, we introduce a complete set of graph coherent states which is much more adapted to the mapping between intertwiners inherent to the action of the gravity Hamiltonian operator.

The article is organized as follows: in the second section we present a summary of the general construction of graph coherent states. In the third section we discuss the gravity Hamiltonian operator(s) in canonical loop quantum gravity, we develop a proposal for the generator and use it to derive the generalized canonical structure, we then define the corresponding graph coherent states which are partially adapted to the gravity Hamiltonian. We close the section with a discussion of certain aspects of the graph change associated to the gravity Hamiltonian operators and their relevance in the construction of the graph coherent states. In the last section, we conclude with a summary of our results and some outlooks.

\section{Graph coherent states in loop quantum theories}\label{sec2}

In this section we review the general construction of the graph coherent states introduced in \cite{GCS1} within the framework of loop quantum theories. The construction is realized on the vertex Hilbert space $\cal{H}_{\text{vtx}}$, obtained from averaging the states in the kinematical Hilbert space $\cal{H}_{\text{kin}}$ of the theory with respect to diffeomorphisms\footnote{In the context of the construction detailed here, in particular with the indistinguishable special loops prescription explained in the following paragraph, we consider the diffeomorphisms to be $C^1$-diffeomorphisms.} which preserve the vertices of the graphs \cite{Hvtx}. The graph coherent states consist of an infinite, but normalized, superposition of $G$-colored networks (called spin networks when $G=SU(2)$) with different graphs. Though originally inspired from a particular dynamics, the construction is purely kinematical in the sense that it is realized on the Hilbert space $\cal{H}_{\text{vtx}}$ independently of the dynamics of the theory under consideration, and in principle it can be applied with various graph changes \cite{GCS1}.

For a clear presentation of the steps and structures involved in the general construction, we proceed with a concrete implementation of a specific graph change which consists of the addition of closed loops at the vertices of the graphs. In particular, we take the graph change proposed for the regularization of the gravity Hamiltonian constraint \cite{AALM, LQGSC}. Namely, the holonomy replacing the curvature of the connection is taken along a closed oriented loop, associated to a pair of edges at a vertex of a preexisting graph, and which does not overlap with any edge of that graph. We call such loops {\it special loops}, and they could be implemented following slightly different prescriptions\footnote{The main property which characterizes different prescriptions is the tangentiality conditions imposed on the edges of the loop at the vertex to which it is attached. In particular, these conditions could be adjusted in order to make the special loops associated to a pair of edges either indistinguishable or entirely distinguishable. We refer the reader to \cite{AALM,LQGSC,GCS1} for further details and discussion.}.
For the moment, we choose to associate such special loop following the prescription which implies that if two edges $e_I$ and $e_J$ have the same tangent vector $\dot e_I$ at a vertex $v$, then given a third independent edge $e_K$, the loops associated to the pairs $(e_Ie_K)$ and $(e_Je_K)$ are diffeomorphically equivalent. This property translates into considering that the tangentiality conditions in the prescription of a special loop to be restricted to a fixed tangentiality order, e.g.\ the first order. With this modification, the special loop prescription guarantees that the added loops, with the same orientation and associated to pairs of edges which belong to the same wedge\footnote{A wedge is a pair of classes of edges at the same vertex of a graph, each class corresponding to a set of edges which have the same tangent vector at the vertex, and this tangent vector is characterizing the class.\label{Wedge&Class}}, are indistinguishable special loops.

Given a colored graph\footnote{By colored graphs we mean diffeomorphism classes of embedded graphs which label the basis states in $\cal{H}_{\text{vtx}}$, and which are characterized by the same non vanishing irreducible representations assigned to the edges, but with no fixed intertwiners.} $\Gamma^{\cal A}$ with a set of vertices $\text{Ver}(\Gamma^{\cal A})$ and no special loops, one can construct a Hilbert space ${\cal H}^{\Gamma^{\cal A}}$, subspace of $\cal{H}_{\text{vtx}}$, spanned by $G$-colored networks associated to all the colored graphs obtained by attaching a finite number of special loops to the vertices of the graph $\Gamma^{\cal A}$. We call the graph $\Gamma^{\cal A}$ the ancestor graph, and we obtain the following decomposition
\begin{align}
{\cal H}_{\text{vtx}}=\overline{\bigoplus_{\Gamma^{\cal A}} {\cal H}^{\Gamma^{\cal A}}} .
\end{align}
Because of the local nature of assigning the special loops, our construction and analysis can be reduced to a single vertex of a given ancestor graph, and the extension to the whole graph is straightforward. Therefore, once given an ancestor graph $\Gamma^{\cal A}$, the only degrees of freedom left are the numbers of loops associated to the wedges at each vertex, and the intertwiners at the vertices. We then can write
\begin{align}\label{Isom}
 {\cal H}^{\Gamma^{\cal A}} \cong \bigotimes_{\ v\in \Gamma^{\cal A}} {\cal H}_v^{\Gamma^{\cal A}} ,
\end{align}
which states that the space ${\cal H}^{\Gamma^{\cal A}}$ is isomorphic to the tensor product of spaces ${\cal H}_v^{\Gamma^{\cal A}}$ each associated to a vertex $v$ of $\Gamma^{\cal A}$. A space ${\cal H}_v^{\Gamma^{\cal A}}$ is spanned by states labelled by a distribution of special loops at the wedges of the ${\Gamma^{\cal A}}$ at the vertex $v$ and an intertwiner (see \cite{GCS1} for more details).

In order to define the graph coherent states, one has to first introduce what we call a canonical structure on the space ${\cal H}^{\Gamma^{\cal A}}$. Namely, a set of closed operators $a_i(v)$ on ${\cal H}^{\Gamma^{\cal A}}$, $i$ being an index in a finite set ${\cal W}_v$ of cardinality $w_v$, associated to the vertices of $v$ of $\Gamma^{\cal A}$ and which satisfy
\begin{align}\label{aCan}
 \forall v,v' \in \text{Ver}(\Gamma^{\cal A}), \forall i\in {\cal W}_v, \forall j\in {\cal W}_{v'},\ [a_i(v),a_j(v')]=[a^{\dag}_i(v),a^{\dag}_j(v')]=0 ,\ [a_i(v), a_j^\dag(v')]=\delta_{vv'}\delta_{ij}\mathbb{I}_{{\cal H}^{\Gamma^{\cal A}}}.
\end{align}

In our previous work \cite{GCS1}, as a concrete case we considered each operator $a_i(v)$ to be associated to a wedge at a vertex $v$. These operators were defined by first introducing an orthonormal basis in each space ${\cal H}_v^{\Gamma^{\cal A}}$ of which the elements are denoted $\ket{\iota_{\{n_i\};\{m_i\}}^\alpha}$, such that $m_i\leq n_i$ with each $n_i$ is the number of special loops associated to the wedge $i$ and $\iota^\alpha$ labels the intertwiner at the vertex, and which satisfy\footnote{For convenience, we drop the label of the vertex $v$ every time we deem it unnecessary or cumbersome.}
\begin{align}
 \forall i\in {\cal W}_v,\ \forall n_i \in \mathbbm{N},\ \forall m_i\leq n_i\ , \qquad a_i^{n_i-m_i} \ket{\iota_{\{n_i\};\{m_i\}}^\alpha} \neq 0\ ,\ a_i^{n_i-m_i+1} \ket{\iota_{\{n_i\};\{m_i\}}^\alpha} = 0\ .
\end{align}
As it will become clear later, the positive integers $m_i$ label the vacuum states selected by the canonical annihilation operators.
The operators $\{a_i,{a}_i^\dag\}$ are then defined through their actions on the states $\ket{\iota_{\{n_i\};\{m_i\}}^\alpha}$ as follows
\begin{subequations}\label{GenAC2}
\begin{alignat}{2}
 \nonumber & \forall k\in {\cal W}_v,\ \forall n_k \in \mathbbm{N},\ &&\forall m_k\leq n_k,\  \\ &\forall \ket{\iota_{\{n_i\};\{m_i\}}^\alpha} \in {\cal H}_v^{\Gamma^{\cal A}} , 
 &&\ a_k \ket{\iota_{\{n_i\};\{m_i\}}^\alpha}=\sqrt{n_k-m_k}\ \ket{\iota_{\{...,n_k-1,...\};\{m_i\}}^\gamma} ,\\ 
 &\ 
 &&\ {a}_k^\dag \ket{\iota_{\{n_i\};\{m_i\}}^\alpha}=\sqrt{n_k-m_k+1}\ \ket{\iota_{\{...,n_k+1,...\};\{m_i\}}^\sigma} .
\end{alignat}
\end{subequations}
We call the operators $a_i$ and $a_i^\dag$ generalized canonical annihilation and creation operators respectively. Equations \eqref{GenAC2} define the operators $\{a_i,{a}_i^\dag\}$ using a choice of mappings between the intertwiners $\iota^\alpha$, and they imply that conditions \eqref{aCan} are satisfied up to conditions on different pairs $\{a_i,{a}_i^\dag\}$ at the same vertex. The remaining conditions constrain the mappings between the intertwiners via
\begin{align}
 \forall i,j\in {\cal W}_v,\ [a_i,a_j]=0\ .
\end{align}
implying that the intertwiner mappings are not independent, but they are also not unique. This means that there is a freedom in choosing the canonical operators $\{a_i,{a}_i^\dag\}$ encoded in the choice of intertwiner mappings.

Note that the states $\ket{\iota_{\{m_i\};\{m_i\}}^\alpha}$ which satisfy
\begin{align}
 \forall i\in {\cal W}_v,\ a_i \ket{\iota_{\{m_i\};\{m_i\}}^\alpha} = 0\ ,
\end{align}
are understood as vacuum states from which arbitrary states $\ket{\iota_{\{n_i\};\{m_i\}}^\alpha}$ are generated via the action of the operators $a_i^\dag$. We denote these vacuum states as $\ket{0_{\{m_i\}}^\alpha}$, and the sub-Hilbert space they span as ${\cal K}_v$, which corresponds to the common kernel of the operators $a_i(v)$.

In general, given a set of commuting, closed, injective\footnote{The injectivity is required on the separable subspaces selected by the graph change where the canonical structure is to be defined. In general, a graph changing operator defined on ${\cal H}_\text{vtx}$ is not injective on its whole domain, however some of its restrictions to the separable subspaces could be. This means that one could proceed with the construction of the canonical structures on the these subspaces where injectivity is satisfied. Finally, note that the issue of injectivity is not related to the graph change only, but also to the intertwiner mapping inherent to the operator under consideration.} and densely defined operators $\{{\cal O}_i\}_{i\in {\cal W}_v}$ which create the same type of graph excitations at a vertex $v$, such as adding a special loop, the general procedure to obtain a corresponding canonical structure is to consider a set of operators $\{\widetilde {\cal O}_i\}_{i\in {\cal W}_s}$ which satisfy
\begin{align}\label{LinIso}
 \forall i\in {\cal W}_v,\ \forall \psi \in {\cal D}({\cal O}_i),\ \frac{\widetilde {\cal O}_i \psi}{\norm{\psi}} = \frac{{\cal O}_i \psi}{\norm{{\cal O}_i \psi}}.
\end{align}
In other words, the operators $\widetilde {\cal O}_i$ are linear isometries which preserve the intertwiner mappings induced by the operators ${\cal O}_i$, that is, two operators ${\cal O}_i$ and $\widetilde {\cal O}_i$ map a given normalized intertwiner to the same normalized intertwiner, with the difference that $\widetilde {\cal O}_i$ preserves the norm, while ${\cal O}_i$ does not necessarily. This step consists of roughly ``normalizing'' the operators ${\cal O}_i$. The operators $\widetilde {\cal O}_i$ are used to define the canonical structure by the identification
\begin{align}
 \forall i\in {\cal W}_v, \ \widetilde {\cal O}_i = a_i^\dag {\cal V}_i \ ,
\end{align}
such that the operator ${\cal V}_i=(a_i a_i^\dag)^{-1/2}$ is diagonal in the colored network basis, and given a basis element $\ket{\iota^\alpha}$ we have
\begin{align}
 {\cal V}_i \ket{\iota^\alpha} = \frac{1}{\sqrt{k_i+1}}\ \ket{\iota^\alpha} ,
\end{align}
where $k_i$ is the difference in the number of graph excitations (e.g.\ special loops) associated to the structure (e.g. wedge) $i$ at the vertex, between the given state and the vacuum state it is generated from. The vacuum states are spin networks chosen as elements of an orthonormal basis which span the kernel of the operators ${\cal O}_i^\dag$.

The association of the generalized canonical operators to the wedges of a graph is an example of how one can construct a canonical structure. As discussed in \cite{GCS1}, this choice is particularly adapted to the action of Yang-Mills Hamiltonian operator in LQG. However, as we will see later, we will introduce another canonical structure which is more adapted to the action of the gravity Hamiltonian operator.

Once we have a canonical structure, we define the graph coherent vertices as being the eigenvectors of the chosen generalized annihilation operators at each vertex, that is
\begin{align}\label{GGCV}
 \forall v\in \Gamma^{\cal A},\quad \forall i\in {\cal W}_v,\quad a_i\ket{Z_v}= z_i\ket{Z_v}\quad,\quad Z_v:=\{ z_i\}\in\mathbb{C}^{w_v}  .
\end{align}
These states are obtained from the vacuum states selected by the canonical structure as
\begin{align}\label{GGCVexp}
 \forall \ket{Z_v},\quad \exists!\ \ket{0_{v,\{m_i\}}^\alpha} \in {\cal K}_v\ :\quad \ket{Z_v} = \prod \limits_{i=1}^{w_v}e^{ z_i a_i^\dag - \bar z_i a_i} \ket{0_{v,\{m_i\}}^\alpha} =: \ket{Z_v, 0_{v,\{m_i\}}^\alpha} .
\end{align}
Finally the graph coherent states are defined as
\begin{align}\label{GGCS}
\ket{Z_{\Gamma^{\cal A}}, 0_{\Gamma^{\cal A}}}:= \bigotimes_{v\in \Gamma^{\cal A}}\ket{Z_v, 0_{v,\{m_i\}}^\alpha} ,
\end{align}
and they are labeled by a colored ancestor graph $\Gamma^{\cal A}$, and to each vertex of $\Gamma^{\cal A}$ is associated a set of complex numbers $Z_v$ representing the eigenvalues of the generalized annihilation operators, and a set of vacuum states $\{0_v^\alpha\}$.

This concludes our overview of the general construction of graph coherent states in loop quantum theories. In the next section, we develop a specific canonical structure on the space $\cal{H}_{\text{vtx}}$ for loop quantum gravity ($G=SU(2)$), which is more adapted to the canonical gravitational quantum dynamics.


\section{Graph coherent states from gravity Hamiltonian}\label{sec3}

\subsection{LQG gravity Hamiltonian}

The gravity Hamiltonian operator in loop quantum gravity can be defined through its action on a (dual) spin network function $\psi_{\Gamma}$ in ${\cal H}_{\text{vtx}}$ as
\begin{align}\label{HamOp}
     H(N) \psi_{\Gamma}= \left(\sum \limits_{ v\in \Gamma} N(v) ( H_v^E +  H_v^L)\right)\psi_{\Gamma} ,
\end{align}
where the Euclidean part operator $H_v^E$ is given by\footnote{In \eqref{EucHam}, we imposed a choice of ordering of the operators, and a choice of symmetrization using the adjoint operators denoted by $^\dag$ understood as the adjoint action on the space ${\cal H}_\text{vtx}$.}
\begin{align}\label{EucHam}
  H_v^E := Q(v) \sum \limits_{I,J} \left( \text{Tr}_N^{( l)}\left[h_{\alpha_{IJ}}  \Upsilon_{IJ} \right]^* + (\text{Tr}_N^{( l)}\left[ h_{\alpha_{IJ}}  \Upsilon_{IJ}  \right]^*)^\dag\right) ,
\end{align}
while the Lorentzian part operator $H_v^L$ could be defined using two different prescriptions: either $H_v^L$ is taken to be proportional to the curvature operator $ R$ introduced in \cite{Curv} and dependent only on the flux operators $ P$, or it is obtained using Thiemann identities which involve the Euclidean part operator, the volume operator $ V$ and the holonomy operators\footnote{Note that the actual holonomy operators in LQG are defined only on the kinematical Hilbert space $\cal{H}_{\text{kin}}$. The term holonomy operator that we use here should be understood as the operator on ${\cal H}_\text{vtx}$ whose action on a dual state, i.e.\ state in $\cal{H}_{\text{vtx}}$, is given by the dual action of the actual holonomy operator on the corresponding kinematical state in $\cal{H}_{\text{kin}}$.} $ h$, for more details see \cite{LQG1,QSD1}.

In \eqref{EucHam}, the capital indices in the ordered sum run through all the edges meeting at the vertex $v$, $\text{Tr}_N^{( l)}$ stands for the normalized trace for the representation\footnote{The construction of the Hamiltonian operator and of the graph coherent states do not depend on the specific choice of the representation of the holonomies, hence it is left arbitrary but assumed to be fixed.} $l$, i.e.\ $\text{Tr}_N^{( l)}[\tau_i^{(l)} \tau_j^{(l)}]=\delta_{ij}$, and the $*$ stands for the dual mapping of the operators from the kinematical Hilbert space to the dual Hilbert space \cite{Hvtx}. The operator $\Upsilon_{IJ}$ could also be defined using two different prescriptions, namely
\begin{align}\label{Yop}
  \Upsilon_{IJ} := \left\{
 \begin{array}{l}
 \sum \limits_K \epsilon^{IJK} [ {V} , h_{s_K}^{-1} ] h_{s_K} \ , \\
 \ \text{or}\\
 \tau_k \left(\epsilon^{ijk} \epsilon_{IJ} P_i^I  P_j^J \right)\widehat{V^{-1}} ,
 \end{array}
\right.
\end{align}
where $\epsilon^{IJK}=-1,0,1,$ depending on the orientation of the triple of edges $(e_I,e_J,e_K)$, $\epsilon_{IJ}$ equals zero when $\dot e_I$ and $\dot e_J$ are collinear, and equals $1$ otherwise. In the second expression the operator $\widehat{V^{-1}}$ is the ``inverse-volume'' operator defined using the Tikhonov regularization \cite{Tikhonov} for the volume operator (see \cite{Length}). Finally, the coefficient $Q(v)$ in \eqref{EucHam} is a factor which depends on the valence of the vertices and on the choice of the operator $\Upsilon_{IJ}$.

The holonomy operators $ h_\alpha$ are the holonomies associated to specific closed loops $\alpha$ at a vertex. For the operator to be defined on ${\cal H}_\text{vtx}$, the added closed loops are chosen to be special loops and could be made either distinguishable or indistinguishable at the wedges. There is also another regularization of the Hamiltonian, the special edges regularization \cite{QSD1}, where the closed loops are partially overlapping with the edges meeting at the vertices of the graph and are completed by a new edge. However, the Hamiltonian operator with such regularization cannot be defined on ${\cal H}_\text{vtx}$, but either on $\cal{H}_{\text{kin}}$ using the so called URST topology \cite{LQG1,RovSmo94}, or on the full diffeomorphism invariant Hilbert space $\cal{H}_{\text{diff}}$ when the lapse $N$ is taken to be a constant\footnote{In this case, one would still have to modify the general expression of the regularized Hamiltonian on the kinematical Hilbert space $\cal{H}_{\text{kin}}$, by including some projectors associated to the ancestor graph, in order to be able to define a densely defined dual operator on $\cal{H}_{\text{diff}}$.}. We further comment on the special edge regularization in the context of our construction of graph coherent states at the end of section \ref{sec3}.

The presence of the holonomy operators $ h_\alpha$ in the expression of $ H_v^E$ makes the operators $ H_v^E$, and consequently $ H(N)$, graph changing operators, i.e.\ they map the graphs they act on to other graphs with a different distribution of special loops at the vertices. 

Naturally, the question is then whether one can use directly the Hamiltonian operator, in particular the part of the Euclidean operator which creates loops, in order to induce a canonical structure or not. The answer is unfortunately negative, for the simple reason that the operators which create loops, namely
\begin{align}\label{NSymEucOp}
 (\text{Tr}_N^{( l)}\left[ h_{\alpha_{IJ}}  \Upsilon_{IJ} \right]^*)^\dag ,
\end{align}
and even their sum as in \eqref{EucHam}, are not injective. This is due to the presence of the volume (or inverse volume) operator in the $\Upsilon_{IJ}$ operators, which has a non trivial kernel for every graph configuration at a vertex. The injectivity of the operator used to generate the canonical structure is essential to construct the canonical operators following the method we propose. In this case, one could think of restricting the construction to the orthogonal complement of the kernel of the volume operator, however it is not clear whether this could work because we do not know if the operator in \eqref{OrigOp} preserves the orthogonal complement of the volume kernel. This is an issue which we do not investigate in this article, and we leave for a future work. We are therefore forced to consider a different operator than the operators \eqref{NSymEucOp}, while still retaining to the maximal extent the structure of the intertwiner mappings present in the gravity Hamiltonian. 

In this article, we propose to use the operators
\begin{align}\label{mEucOp}
 \left(\text{Tr}_N^{( l)}\left[ h_{\alpha_{IJ}} \tau_k  Y_{IJ}^k \right]^* \right)^\dag ,
\end{align}
where
\begin{align}
  Y_{IJ}^k := \epsilon^{ijk} \epsilon_{IJ} P_i^I  P_j^J ,
\end{align}
One can see that the difference between the operators in \eqref{mEucOp} and the ones in \eqref{EucHam}, when $\Upsilon_{IJ}$ are taken to be the second operators in \eqref{Yop}, is the absence of the inverse volume operator. In fact, the operators \eqref{mEucOp} form the Euclidean part of the physical Hamiltonian in the context of the loop deparametrized theory with respect to a massless scalar field \cite{DGKL, GieThiem12, AALM}. As it is shown in the appendix \ref{appendix}, the operators \eqref{mEucOp} are injective operators, except on the specific degenerate subspaces corresponding to vertices which have no more than two collinear classes of edges (see footnote \ref{Wedge&Class}), We call such vertices degenerate vertices. Since the operators \eqref{mEucOp} do not generate such vertices, these are simply excluded from the construction.

\subsection{Graph coherent states with special loops}

We come now to the construction of graph coherent states which are partially adapted to the gravity Hamiltonian in LQG. By partially adapted we mean that the graph coherent states will implement the graph change induced by the action of the Hamiltonian operator, but they will incorporate only partially the mapping between intertwiners, which is induced by the Euclidean part in \eqref{EucHam}. We therefore focus on the operators
\begin{align}\label{NotNormOp}
 \left(\text{Tr}_N^{( l)}\left[ h_{\alpha_{IJ}} \tau_k  Y_{IJ}^k \right]^* \right)^\dag ,
\end{align}
which are considered to be the operators creating special loops, while their adjoint operators are removing them. An operator as in \eqref{NotNormOp} associates (diffeomorphically equivalent) special loops to the wedge $(IJ)$. Hence one would first consider to use the intertwiner mapping induced by the operators in \eqref{NotNormOp} in order to build a canonical structure \eqref{aCan}, where each pair of canonical operators is associated to a wedge of an ancestor graph. However, it is easy to observe that at a given vertex $v$, the operators \eqref{NotNormOp} associated to the wedges at $v$ do not commute. This implies that with these intertwiner mappings, the pairs of canonical operators associated to different wedges at the same vertex would not always commute with each other, and hence one would fail to have a canonical structure at the vertex. Thus, we are obliged to look for a different way to implement these intertwiner mappings.

Since we are interested in the Hamiltonian operator as a whole, the idea is to take the sum of the operators in \eqref{NotNormOp} associated to the same vertex, as a base to construct the canonical structure. This means that instead of looking at the wedges separately, and associating a pair of canonical operators to each one of them, we will try to build only one pair of canonical operators associated to the vertex itself. Such structure would trivially solve the problem of commutation since there would be only one canonical pair per vertex.
The closed, densely defined, and injective \footnote{The proof of these properties is presented in the appendix \ref{appendix}. Note that, as mentioned earlier, the injectivity is required only on the Hilbert spaces ${\cal H}^{\Gamma^{\cal A}}$ associated to ancestor graphs where the vertex $v$ is not degenerate.} operator ${\cal C}_v$ that we consider to generate the canonical structure is then
\begin{align}\label{OrigOp}
 {\cal C}_v := \sum \limits_{I,J} \left(\text{Tr}_N^{( l)}\left[ h_{\alpha_{IJ}} \tau_k  Y_{IJ}^k \right]^* \right)^\dag .
\end{align}
As we mentioned earlier, this operator contributes in forming the physical Hamiltonian in the deparametrized model with respect to a massless scalar field. More precisely, it corresponds to the Euclidean part of the square of the physical Hamiltonian operator \cite{DGKL, AALM}. From now on, we refer to the operator ${\cal C}_v$ as the non symmetric Euclidean operator.
Using \eqref{LinIso}, ${\cal C}_v$ defines the linear isometry $\widetilde {\cal C}_v$, which naturally induces a canonical structure through the identification
\begin{align}\label{CanStr}
 a_v^\dag {\cal V}_v = \widetilde {\cal C}_v\ ,\quad \text{such that}\ {\cal V}_v:=(a_v a_v^\dag)^{-1/2} ,
\end{align}
where $a_v$ and $a_v^\dag$ are the induced annihilation and creation operators respectively. To understand the details of the construction, let us elaborate on the canonical structure defined in \eqref{CanStr}. Given an ancestor graph ${\Gamma^{\cal A}}$, we first note that we can introduce a decomposition of the Hilbert space ${\cal H}_v^{\Gamma^{\cal A}}$ associated to a vertex $v$, namely
\begin{align}
 {\cal H}_v^{\Gamma^{\cal A}} = \overline{\bigoplus_n {\cal H}_v^n} ,
\end{align}
where a space ${\cal H}_v^n$ is the finite dimensional Hilbert space of states with $n$ special loops at the vertex $v$. We denote the normalized states in ${\cal H}_v^n$ which belong to the kernel of ${\cal C}_v^\dag$ as $\ket{0_n^\alpha}$, these are the vacuum states in our construction. The successive action of $ {\cal C}_v$ on the vacuum states generates states which, once normalized, correspond to the number states $\ket{n;0_m^\alpha}$, namely
\begin{align}\label{NumSts}
 \ket{n;0_m^\alpha}:= (\widetilde {\cal C}_v)^{n-m} \ket{0_m^\alpha} = \frac{1}{\norm{({\cal C}_v)^{n-m} \ket{0_m^\alpha}}}({\cal C}_v)^{n-m} \ket{0_m^\alpha} ,
\end{align}
where $n$ denotes the total number of special loops in the state, while $m$ denotes the number of special loops in the vacuum state they come from. Each number state $\ket{n;0_m^\alpha}$ is a superposition of dual spin network states (at the vertex $v$) associated  to graphs representing all possible distributions of $n-m$ loops at the wedges of the vacuum state with $m$ loops.
Note that for the vacuum states, we use the short notation $\ket{0_m^\alpha}$ which replaces $\ket{m; 0_m^\alpha}$. These number states are the eigenstates of the number operator ${\cal N}_v:= a_v^\dag a_v$, and we have
\begin{subequations}\label{CanOpGr}
\begin{align}
 \forall n \in \mathbbm{N},\ \forall m \leq n,\ \forall \ket{n;0_m^\alpha} \in {\cal H}_v^n , 
 \qquad a_v \ket{n;0_m^\alpha} &= \sqrt{n-m}\ \ket{n-1;0_m^\alpha} ,\\ 
 \qquad a_v^\dag \ket{n;0_m^\alpha} &= \sqrt{n-m+1}\ \ket{n+1;0_m^\alpha} ,\\
 \qquad {\cal N}_v \ket{n;0_m^\alpha} &= (n-m)\ \ket{n;0_m^\alpha} .
\end{align}
\end{subequations}

At this point, one can define the graph coherent states obtained from a selected set of vacuum states $\{\ket{0_{v,m}^\alpha}\}$, and the canonical operators $(a_v,a_v^\dag)$ induced by the operator $ {\cal C}_v$. Similarly to \eqref{GGCV}, one first defines the graph coherent vertices as the eigenvectors of the annihilation operators
\begin{align}\label{GGCVgr}
 \forall v\in \Gamma^{\cal A},\quad a_v\ket{z_v}= z_v\ket{z_v}\quad,\quad z_v\in\mathbb{C} ,
\end{align}
and are given as
\begin{align}\label{GGCVgrexp}
 \forall \ket{z_v},\quad \exists!\ \ket{0_{m,v}^\alpha} \in {\cal K}_v\ :\quad \ket{z_v} = e^{ z_v a_v^\dag - \bar z_v a_v} \ket{0_{m,v}^\alpha} =: \ket{z_v, 0_{m,v}^\alpha} .
\end{align}
The graph coherent states induced by the operator $ {\cal C}_v$ with an ancestor graph ${\Gamma^{\cal A}}$ are then defined as
\begin{align}\label{GGCSgr}
\ket{Z_{\Gamma^{\cal A}}, 0_{\Gamma^{\cal A}}}:= \bigotimes_{v\in \Gamma^{\cal A}}\ket{z_v, 0_{m,v}^\alpha} .
\end{align}
This concludes the construction of the graph coherent states associated to the generalized canonical structure induced by the non symmetric Euclidean operator $ {\cal C}_v$.

Let us now make some comments about the construction of a canonical structure when a different graph change for the Hamiltonian is considered. As mentioned earlier, one can consider different regularizations for the Hamiltonian, inducing an Euclidean operator which generates different graph changes, namely the indistinguishable special loop, the distinguishable special loop, or the special edge. The indistinguishable special loops corresponds to the graph change we considered in the construction above. From the framework detailed above, one can deduce that the construction can also be realized with the distinguishable special loops, with no difference with respect to the indistinguishable special loops case. This is mainly because the non symmetric Euclidean operator ${\cal C}_v$ at a vertex remains an injective operator when considering distinguishable special loops. However, in the case of the special edge prescription, if we consider the proper operator ${\cal C}_v$ whether defined on $\cal{H}_{\text{kin}}$ or $\cal{H}_{\text{diff}}$, the coupling induced by the overlapping holonomies generates a much more complex structure at the vertex, and we do not know whether the injectivity would be satisfied. In particular, for certain ancestor graphs, the repetitive action of the ${\cal C}_v$ would produce components where the edges at the vertex get annihilated once the representations associated to them vanish. Some of such components could eventually lead to a saturated structure, where the vertex itself is annihilated. At the moment, it is not yet clear how to deal with such components in our framework. We leave this question for future studies. 

Finally, note that since the graph coherent states are by construction canonical coherent states, they satisfy the standard coherence and peakedness properties with respect to the canonical operators $a_v$ and $a_v^\dag$. The further interesting aspects to look into would be the coherence properties with respect to specific operators of interest. One such operator would be the Euclidean part of the gravity Hamiltonian, in particular the operators ${\cal C}_v$ and their adjoints ${\cal C}_v^\dag$. We will present the results of the analysis of the coherence properties with respect to these operators and their combination in a follow-up article.


\section{Summary \& outlooks}

In this article, we reviewed briefly the general construction of graph coherent states in loop quantum theories introduced in \cite{GCS1}, and we further constructed a family of graph coherent states adapted to the gravity Hamiltonian operator, regularized following the special loops prescription. We approached the construction of these graph coherent states from the perspective of using the non symmetric Euclidean operator as a generator of the generalized canonical structure on the vertex Hilbert space ${\cal H}_\text{vtx}$. 

We first established that for the vacuum gravity constraint operator, the construction may not go through because of the presence of the volume operator (or the inverse volume operator), which makes the Euclidean operator not injective. In this case, the condition in order to realize the construction on the relevant subspaces is for the Euclidean operator to preserve the orthogonal complement of the kernel of the volume operator. However, we so far did not establish the validity or not of this statement. We hence moved to considering a modified Euclidean operator ${\cal C}_v$, which retains part of the intertwiner mappings of the Euclidean operator in the general Hamiltonian constraint, and which is also present in the physical Hamiltonian obtained in the context of the deparametrized model with respect to a massless scalar field. Since ${\cal C}_v$ is injective on the relevant subspaces of ${\cal H}_\text{vtx}$, we were able to introduce a generalized canonical structure, associating a pair of canonical operators to each (non degenerate) vertex of a given graph, which retains fully the intertwiner mapping induced by the operator ${\cal C}_v$. Consequently, we defined the new family of graph coherent states for gravity compatible with the gravity Hamiltonian operator.

The coherence properties of this new family of graph coherent states with respect to the non symmetric Euclidean operator ${\cal C}_v$, in addition to other operators of interest such the symmetric Euclidean and Lorentzian parts of the Hamiltonian constraint, are currently under investigation. The results will be presented in a follow-up article.

Finally, it would be of great interest to further explore the role and potential of such graph coherent states in the context of constructing a semi-classical limit for graph changing operators in general, and in the analysis of the dynamics generated by the graph changing Hamiltonian operators in particular. We leave these questions for future research.


\section*{Acknowledgements}
The author thanks Ilkka M\"akinen for useful comments. This work was supported by the Polish National Science Center OPUS 15 Grant nr 2018/29/B/ST2/01250 and by the project BA 4966/1-1 of the German Research Foundation (DFG).

\appendix

\section{Proof of the properties of the operator ${\cal C}_v$ with special loops}\label{appendix}

\setlength{\parindent}{0pt}

We present the proofs that the operator ${\cal C}_v$ defined in \eqref{OrigOp}, namely
\begin{align}
 {\cal C}_v = \sum \limits_{I,J} \left(\text{Tr}_N^{( l)}\left[ h_{\alpha_{IJ}} \tau_k  Y_{IJ}^k \right]^* \right)^\dag ,
\end{align}
is closed, densely defined on ${\cal H}_\text{vtx}$, and its restrictions are injective on the Hilbert spaces ${\cal H}^{\Gamma^{\cal A}}$ associated to the ancestor graphs ${\Gamma^{\cal A}}$ where the vertex $v$ is not degenerate. \\

We denote by $(C_v^E)_\epsilon$ the operator
\begin{align}
 (C_v^E)_\epsilon := \sum \limits_{I,J} \text{Tr}_N^{( l)}\left[ h_{\alpha_{IJ}}^\epsilon \tau_k  Y_{IJ}^k \right] ,
\end{align}
defined on the kinematical Hilbert space ${\cal H}_\text{kin}$, where $\epsilon$ stands for the coordinate size of the special loops $\alpha_{IJ}$. The dependence on this coordinate size would be eventually removed by taking the limit $\epsilon \rightarrow 0$ on the Hilbert space ${\cal H}_\text{vtx}$. The operator $(C_v^E)_\epsilon$ is constructed by regularization methods using cylindrical functions, i.e.\ a spin network function. This means that the action of $(C_v^E)_\epsilon$ is defined on all the spin network functions. It follows that the domain ${\cal D}$ of $(C_v^E)_\epsilon$ is taken to be the linear span of the spin network functions $Cyl$, which we take to be the inner product space of which the completion is ${\cal H}_\text{kin}$. Hence, the operator $(C_v^E)_\epsilon$ is densely defined by construction, and its image ${\cal R}$ is a subspace of $Cyl$. The inner product space $Cyl$ can be decomposed into a direct sum of inner product spaces $Cyl_{\Gamma^{\cal A}}$, 
\begin{align}\label{DecompCyl}
 Cyl = \bigoplus_{\Gamma^{\cal A}} Cyl_{\Gamma^{\cal A}} ,
\end{align}
where each $Cyl_{\Gamma^{\cal A}}$ is the span of spin network states with graphs consisting of an ancestor graph $\Gamma^{\cal A}$ with arbitrary numbers of special loops at the vertices.
We then can introduce the algebraic dual $Cyl^*$ of $Cyl$ which, following from \eqref{DecompCyl}, decomposes as
\begin{align}\label{DecompCyl*}
 Cyl^* \cong \bigoplus_{\Gamma^{\cal A}} Cyl_{\Gamma^{\cal A}}^*\ ,
\end{align}
and on which we can define the dual (transpose) operator $(C_v^E)_\epsilon^*$ as
\begin{align}
 (C_v^E)_\epsilon^* \ :\ {\cal D}_* \subset Cyl^* \longrightarrow Cyl^* \ |\ 
  & {\cal D}_* := \{ \chi \in Cyl^* : \exists c \geq 0 : \forall \psi \in {\cal D} , \abs{\chi \left( (C_v^E)_\epsilon \psi \right)} \leq c \norm{\psi}\} , \notag \\
 & \forall \chi \in {\cal D}_*, \forall \psi \in {\cal D} : (C_v^E)_\epsilon^* \chi \left( \psi \right) = \chi \left( (C_v^E)_\epsilon \psi \right) ,
\end{align}
and we write
\begin{align}
 (C_v^E)_\epsilon^* = \sum \limits_{I,J} \text{Tr}_N^{( l)}\left[ h_{\alpha_{IJ}}^\epsilon \tau_k  Y_{IJ}^k \right]^* .
\end{align}
Using the definition of the domain ${\cal D}_*$, one can easily show that every dual spin network function (every spin network function in $Cyl$ defines a linear functional in $Cyl^*$, through the inner product on $Cyl$, which we call the dual spin network) belongs to ${\cal D}_*$. This means that ${\cal D}_*$ is dense in $Cyl^*$, hence $(C_v^E)_\epsilon^*$ is densely defined on $Cyl^*$.

We come now to the implementation of diffeomorphism invariance and the construction of ${\cal H}_\text{vtx}$. The Hilbert space ${\cal H}_\text{vtx}$ is obtained by averaging the states in each $Cyl_{\Gamma^{\cal A}}$ with respect to diffeomorphisms which preserve $\text{Ver}(\Gamma^{\cal A})$ using a rigging map $\eta$. The images $\eta\left( Cyl_{\Gamma^{\cal A}}\right)=:Cyl_{\text{inv},\Gamma^{\cal A}}^*$ are subspaces of $Cyl^*$. Introducing the space $Cyl_{\text{inv}}^*$, 
\begin{align}\label{DecompCyl*inv}
 Cyl_\text{inv}^* := \bigoplus_{\Gamma^{\cal A}} Cyl_{\text{inv},\Gamma^{\cal A}}^*\ ,
\end{align}
the vertex Hilbert space ${\cal H}_\text{vtx}$ is defined as
\begin{align}\label{DecompHvtx}
 {\cal H}_\text{vtx} := \overline{\bigoplus_{\Gamma^{\cal A}} Cyl_{\text{inv},\Gamma^{\cal A}}^*}\ ,
\end{align}
where the completion is taken with respect to the inner product induced from the action of elements of $Cyl^*$ on elements of $Cyl$, namely
\begin{align}
 \forall, \chi, \psi \in Cyl,\ (\eta(\chi),\eta(\psi)) := \eta(\chi)(\psi) .
\end{align}

One then can show that the limit $\lim \limits_{\epsilon\rightarrow 0} (C_v^E)_\epsilon^*$ exists on ${\cal H}_\text{vtx}$ \cite{Hvtx}, 
\begin{align}
 \lim \limits_{\epsilon\rightarrow 0} (C_v^E)_\epsilon^* = (C_v^E)^* = \sum \limits_{I,J} \text{Tr}_N^{( l)}\left[ h_{\alpha_{IJ}} \tau_k  Y_{IJ}^k \right]^* ,
\end{align}
and it maps each $Cyl_{\text{inv},\Gamma^{\cal A}}^*$ to a subspace of $Cyl_{\text{inv},\Gamma^{\cal A}}^*$. Since $(C_v^E)_\epsilon^*$ is densely defined on $Cyl^*$, and denoting ${\cal D}_\eta := {\cal D}_* \cap {\cal H}_\text{vtx}$, it follows that\\

\hspace{1cm} {\it $\bullet$ Proposition 1: $(C_v^E)^* : {\cal D}_\eta \rightarrow {\cal H}_\text{vtx}$ is densely defined on ${\cal H}_\text{vtx}$, and its adjoint $\left[(C_v^E)^*\right]^\dag =: {\cal C}_v$ is a closed operator on ${\cal H}_\text{vtx}$.}\\

Moreover, from the definition of the adjoint one can show that its domain ${\cal D}_\eta^\dag$ contains every (partial) diffeomorphism invariant spin network function, hence \\

\hspace{1cm} {\it $\bullet$ Proposition 2: ${\cal C}_v : {\cal D}_\eta^\dag \subset {\cal H}_\text{vtx} \rightarrow {\cal H}_\text{vtx}$ is densely defined on ${\cal H}_\text{vtx}$, and $(C_v^E)^*$ is closable.}\\

Note that in this article, we assume that every closable operator is replaced by its closure. 

Going back to the operator $(C_v^E)_\epsilon$, one can compute the action of each operator $ \text{Tr}_N^{( l)}\left[ h_{\alpha_{IJ}}^\epsilon \tau_k  Y_{IJ}^k \right]$ on a spin network function \cite{AALM, ALM}, and using that, one can show that the only spin network states in the kernel ${\cal K}_v^\epsilon$ of $(C_v^E)_\epsilon$ correspond to states where the vertex $v$ is degenerate, i.e.\ the vertex has no more than two collinear classes of edges (see footnote \ref{Wedge&Class} for the definition of a class of edges). Since such states do not belong to the image of $(C_v^E)_\epsilon$, one can establish that the restrictions of the operator $(C_v^E)_\epsilon$ to each $Cyl_{\Gamma^{\cal A}}$, where the vertex $v$ in ${\Gamma^{\cal A}}$ is not degenerate, are injective operators on their respective domains $Cyl_{\Gamma^{\cal A}}$.

The injectivity of the restrictions of $(C_v^E)_\epsilon$ on the specific spaces $Cyl_{\Gamma^{\cal A}}$ implies that $(C_v^E)_\epsilon^* \left(Cyl_{\Gamma^{\cal A}}^* \right)$, the image of $Cyl_{\Gamma^{\cal A}}^*$ by $(C_v^E)_\epsilon^*$, is dense in $Cyl_{\text{inv},\Gamma^{\cal A}}^*$ and consequently it is dense in its completion ${\cal H}^{\Gamma^{\cal A}} := \overline{Cyl_{\text{inv},\Gamma^{\cal A}}^*}$. Hence $(C_v^E)^* (Cyl_{\text{inv},\Gamma^{\cal A}}^* )$ is also dense in ${\cal H}^{\Gamma^{\cal A}}$.
Using the closed range theorem, it follows that\\

\hspace{1cm} {\it $\bullet$ Proposition 3: For every $\Gamma^{\cal A}$ where $v\in \text{Ver}(\Gamma^{\cal A})$ is not a degenerate vertex, the maps
 \begin{align}
 {\cal C}_v : {\cal D}_\eta^\dag \cap {\cal H}^{\Gamma^{\cal A}} \rightarrow {\cal H}^{\Gamma^{\cal A}} \notag ,
 \end{align}
are injective operators on their domains ${\cal D}_\eta^\dag \cap {\cal H}^{\Gamma^{\cal A}}$.} \\

This concludes the proofs of the properties of the operator ${\cal C}_v$.


\bibliography{references}
\bibliographystyle{ieeetr}

\end{document}